\documentclass[%
superscriptaddress,
 amsmath,amssymb,
 aps, twocolumn,
floatfix,
]{revtex4-2}

\usepackage{graphicx}
\usepackage{dcolumn}
\usepackage{bm}
\usepackage[normalem]{ulem}
\usepackage{xcolor} 
\usepackage{float}

\usepackage{enumitem}

\let\centeredsection\section

\usepackage{etoolbox}
\patchcmd{\section}{\centering}{}{}{}

\let\flushleftsection\section
\newcommand{\sectionscenter}{\let\section\centeredsection}
\newcommand{\sectionsleft}{\let\section\flushleftsection}

\let\centeredsubsection\subsection

\usepackage{etoolbox}
\patchcmd{\subsection}{\centering}{}{}{}

\let\flushleftsubsection\subsection
\newcommand{\subsectionscenter}{\let\subsection\centeredsubsection}
\newcommand{\subsectionsleft}{\let\subsection\flushleftsubsection}

\let\centeredsubsubsection\subsubsection

\usepackage{etoolbox}
\patchcmd{\subsubsection}{\centering}{}{}{}

\let\flushleftsubsubsection\subsubsection
\newcommand{\subsubsectionscenter}{\let\subsubsection\centeredsubsubsection}
\newcommand{\subsubsectionsleft}{\let\subsubsection\flushleftsubsubsection}

\begin{document}

\preprint{revtex4.2 preprint}

\title{Searching for structural predictors of plasticity in dense active packings}

\author{Julia A. Giannini}
\email{jagianni@syr.edu}
\affiliation{Department of Physics, Syracuse University, Syracuse, New York 13244, USA}
\affiliation{BioInspired Institute, Syracuse University, Syracuse, New York 13244, USA}
\author{Ethan M. Stanifer}
\affiliation{Department of Physics, University of Michigan, Ann Arbor, Michigan 48109, USA}
 \email{mmanning@syr.edu}
 \author{M. Lisa Manning}
\affiliation{Department of Physics, Syracuse University, Syracuse, New York 13244, USA}
\affiliation{BioInspired Institute, Syracuse University, Syracuse, New York 13244, USA}%

\date{\today}

\begin{abstract}
 In amorphous solids subject to shear or thermal excitation, so-called structural indicators have been developed that predict locations of future plasticity or particle rearrangements. An open question is whether similar tools can be used in dense active materials, but a challenge is that under most circumstances, active systems do not possess well-defined solid reference configurations. We develop a computational model for a dense active crowd attracted to a point of interest, which does permit a mechanically stable reference state in the limit of infinitely persistent motion. Previous work on a similar system suggested that the collective motion of crowds could be predicted by inverting a matrix of time-averaged two-particle correlation functions. Seeking a first-principles understanding of this result, we demonstrate that this active matter system maps directly onto a granular packing in the presence of an external potential, and extend an existing structural indicator based on linear response to predict plasticity in the presence of noisy dynamics. We find that the strong pressure gradient necessitated by the directed activity, as well as a self-generated free boundary, strongly impact the linear response of the system. In low-pressure regions the linear-response-based indicator is predictive, but it does not work well in the high-pressure interior of our active packings. Our findings motivate and inform future work that could better formulate structure-dynamics predictions in systems with strong pressure gradients.
\end{abstract}

\maketitle


\section{\label{sec:intro}Introduction}

Dense amorphous solids -- including powders, granular systems, foams, structural glasses, and colloidal assemblies -- are ubiquitous in nature \cite{berthier_theoretical_2011,weeks_introduction_2017,weeks_properties_2002}. These materials exhibit unique mechanical and dynamic features that emanate from their disordered structure \cite{lerner_micromechanics_2016, schoenholz_structural_2016}. Similarly, in some cases, active matter comprised of self-propelled agents remains disordered as it achieves very high densities; examples of such systems include bacterial assemblies \cite{pierce_hydrodynamic_2018}, cellular tissues \cite{schotz_glassy_2013, bi_motility-driven_2016}, and groups of animals \cite{bottinelli_when_2019,cavagna_empirical_2010}. Although active matter is relatively well-studied at low and intermediate densities \cite{marchetti_hydrodynamics_2013,cates_active_2019}, an important open question is whether the emergent mechanical properties of dense active matter are similar to, or different from, their non-active counterparts \cite{mandal_extreme_2020,berthier_how_2017,berthier_glassy_2019,marchetti_minimal_2016}.

One starting point for answering this question is to analyze properties of inherent or reference states of the amorphous solid that underlies a given dense active material \cite{henkes_extracting_2012, henkes_dense_2020, henkes_dynamical_2011, szamel_long-ranged_2021}. In this framework, one considers how structural information from a static snapshot of the system, usually the positions and sizes of individual particles and the potential energy with which they interact, can provide insight into dynamic yielding behavior when the system is subject to external deformation or activity \cite{falk_dynamics_1998, morse_direct_2021, agoritsas_mean-field_2021, richard_simple_2021}. A large body of work explores structure-dynamics predictions in sheared, athermal disordered solids. In a recent article (Ref.~\citenum{richard_predicting_2020}), Richard \textit{et.~al.}~compare the performance of several classes of structural indicators in identifying localized instabilities or defects in computer glasses which forecast plastic rearrangements under shear strain.

\begin{figure}[ht]
\includegraphics[width = 0.97\linewidth]{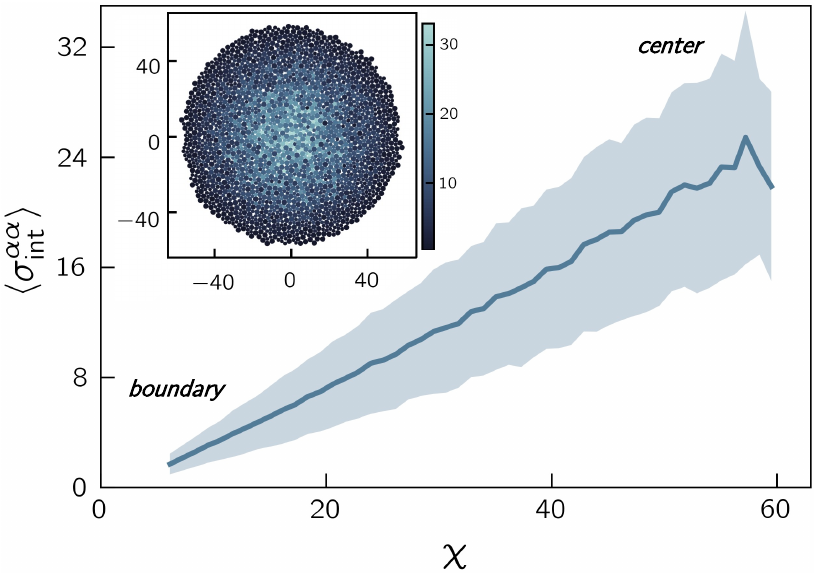}
\caption{\label{fig:intro}  Properties of dense packings of active particles directed towards a central point of interest. At mechanical equilibrium, these packings feature a gradient in interaction pressure that has azimuthal symmetry. The main panel shows the mean per-particle interaction pressure as a function of radius from the edge of the system for an ensemble of $25$ packings with $N = 2048$ and $v_0 = 0.5$. The shaded region shows the standard deviation of the interaction pressure at different locations in the packings. The inset shows an example $N=2048, v_0=0.5$ system with particles colored by the magnitude of their interaction pressures.}
\end{figure}

In this work, we focus on linear-response-based structural metrics, which utilize the spectrum of vibrational modes of a solid computed in the harmonic approximation of the total potential energy. As shown in Ref.~\citenum{richard_predicting_2020} and other works, these metrics are surprisingly good at identifying soft spots, or localized microstructural instabilities, in sheared amorphous solids \cite{manning_vibrational_2011,tong_revealing_2019,richard_universality_2020,wang_low-frequency_2019}. A primary goal of our work is to extend this class of structural indicators to active solids. Thus, a first challenge is to identify an active material with a time-invariant, well-defined reference state, as most active systems are ``self-shearing" and not mechanically stable \cite{morse_direct_2021, briand_spontaneously_2018}. Here, we consider assemblies of active particles that are infinitely persistent in a radial direction towards a central point of interest. As we will show, the symmetry of this biased activity permits a force-balanced steady state, and allows us to exactly map the relevant non-Hamiltonian self-propulsion forces onto an effective external potential. This choice also necessarily introduces a strong interaction pressure gradient and a self-generated free boundary as depicted in Fig.~\ref{fig:intro}. 

Previously, a similar geometry and set of dynamical equations was studied by Bottinelli and Silverberg in a computational model for dense human crowds \cite{bottinelli_emergent_2016, bottinelli_how_2017}. Their study sought to predict density waves or localized excitations that are thought to correspond to dangerous collective behaviors such as trampling or crowd-crush events. Predicting these phenomena from basic structural information or dynamics is an important first step toward avoiding or controlling crowd disasters. Toward this goal, the authors adopted techniques that have previously been deployed in colloidal systems to estimate the system's linear response~\cite{chen_low-frequency_2010, chen_measurement_2011, kaya_normal_2010}, where the dynamical matrix is estimated from long-time averages of two-particle correlation functions. In addition to analyzing particle trajectories from simulated crowds, the same authors applied these techniques to video footage of real human crowds and were indeed able to predict wave-like collective motion, albeit over a very short time window \cite{bottinelli_can_2018,bottinelli_when_2019}.

A significant challenge associated with this framework, which approximates the linear response of the system, is that the equivalence between the dynamical matrix and two-time correlation functions only holds under certain assumptions: namely, that i) the correlation functions are averaged over long time intervals; ii) the dynamics of the system are thermal; and iii) there are no changes to the underlying contact network during the relevant time intervals. In real crowds or self-propelled particle models, none of these assumptions hold. Therefore, the appropriate analogue of the dynamical matrix in systems whose microscopic details are non-Hamiltonian remains unclear.

Our work is also informed by previous research on thin films and other materials with free surfaces, as we expect that the free boundary alone might alter the mobility or linear response of a disordered packing. For example, a study by Sussman \textit{et.~al.}~\cite{msussman_disordered_2015} examines the vibrational modes of unstressed spring networks derived from partially periodic jammed packings with free boundaries, and finds a population of low frequency modes that exhibit an exponential decay in magnitude away from the edges. In contrast, distinct work by Sussman and collaborators~\cite{sussman_disconnecting_2017} finds that there is a decoupling between structure and dynamics near the edge of glassy thin films, where an attractive interaction generates the free boundary. Specifically, the authors use a machine learning approach to show that there are no special structural features near the edge of the material, even though the mobility is higher there. Taken together, this suggests that there may be some material-dependent subtleties in whether the structure and vibrational properties of a solid predict particle rearrangements near a free boundary.

Here, we build the beginnings of a framework for predicting localized rearrangements in dense active matter. We first demonstrate that ``point-of-interest" model systems have well-defined solid reference states, which allow us to map the active forces onto an effective potential that can be encoded in an augmented Hessian or dynamical matrix. Next, we add noisy dynamics to the system to perturb it away from its reference state, and study whether the vibrational spectrum can be used to predict changes in structure. We find that the strong pressure gradients in the system may limit the predictive power of this extended linear response. Ultimately, our results highlight that more sophisticated methods such as non-linear-response-based structural metrics may be required to identify the microstructural entities that determine the stability of active packings.


\section{\label{sec:methods}Methods}

\subsection{Model}

We study an active particle model in two dimensions with overdamped dynamics. Stable packings of $N$ discs are formed by evolving the following single-particle equation of motion from a randomized initial state until force balance is reached: 
\begin{equation}\label{eq:static_eom}
    \dot{\vec{r_i}} = \frac{1}{\Gamma} \vec{F}_{i, \text{int}} +  v_0 \hat{n}_i.
\end{equation} 
Here, $\vec{r_i}$ contains the positional degrees of freedom of particle $i$, $\Gamma$ is a viscous damping coefficient set to unity, $\vec{F}_{i, \text{int}}$ is the net interaction force on particle $i$ by its neighbors, $v_0$ is the magnitude of the self-propulsion velocity (which is the same for all particles), $\hat{n}_i$ is a unit vector pointing in the direction of self propulsion, and $\dot{(\cdot)}$ denotes a time derivative. Pairwise repulsive forces between the particles are determined via the Hertzian soft sphere potential: 
\begin{equation}
    F_{ij, \text{int}}(r_{ij}) =
    \begin{cases}
    \frac{k}{R_{ij}} \left( 1 - \frac{r_{ij}}{R_{ij}}\right)^{3/2} \quad \text{ if } r_{ij} < R_{ij} \\ 
    0 \qquad \text{  else},
    \end{cases}
\end{equation}
where $k$ is the interaction stiffness constant, $R_{ij} = R_i + R_j$ is the sum of the radii of particles $i$ and $j$, and $r_{ij} = \vert \vec{r_j} - \vec{r_i} \vert$ is the distance between $i$ and $j$. We employ 50:50 binary mixtures with a 1:1.4 ratio between the small and large particle radii to discourage crystallization. The direction of the force $\vec{F}_{ij,\text{int}}$ exerted on particle $i$ by $j$ is parallel to the line that connects $j$'s center to $i$'s. In the limit of infinite persistence, in which we form initial reference configurations, $\hat{n}_i$ always points toward a ``point of interest" located at the origin in the plane.

In the presence of translational noise, in which we examine active dynamics initialized from each static reference configuration, the single-particle equation of motion is given by
\begin{equation}\label{eq:noisy_eom}
    \dot{\vec{r_i}} = \frac{1}{\Gamma} \vec{F}_{i, \text{int}} +  v_0 \hat{n}_i + \vec{\eta_i},
\end{equation} 
where $\vec{\eta_i}$ is white noise with zero mean and magnitude $\sigma$. We study dynamics with different levels of noise by examining simulations at different temperatures $T = \frac{\sigma^2 \Gamma}{2}$. The stochastic differential equations (Eq.~\ref{eq:noisy_eom}) are integrated via the velocity Verlet algorithm with a stable timestep determined by examining the relative magnitudes of typical interparticle and self propulsion forces. See Appendix~\ref{sec:appA_model} for more details of our implementation. An example movie of the formation of a static reference configuration and ensuing thermal dynamics is available in the supplemental material.

The dynamics of the polarization direction $\hat{n}_i = \left(\cos{(\theta_i)}, \sin{(\theta_i)} \right)$ of particle $i$ in our noisy simulations are governed by the angular equation of motion,
\begin{equation}\label{eq:angle_eom}
    \dot{\theta_i} = \frac{1}{\tau} \Delta \theta_i,
\end{equation}
where $\Delta \theta_i$ is the (smallest) angular distance between $\hat{n}_i$ and the vector that points from $i$'s center to the point of interest in a given simulation time step, and $\tau$ is a characteristic turning time set to unity.

In the discussion that follows, we present results from an ensemble of static packings with $N \in \left\{256, 512, 1024, 2048, 4096 \right\}$ and $v_0 \in \left\{0.25, 0.5, 1.0 \right\}$, and corresponding noisy dynamics with $T \in \left\{ 0.125, 0.142, 0.165, 0.197, 0.244, 0.320 \right\}$. For each state point, we consider 25 duplicate simulations. Via a simple toy model which we describe in the next section and in Appendix~\ref{sec:appB_toy}, we choose the parameter $k$ such that the maximum packing fraction in the largest configurations does not exceed approximately $\phi \sim 1.3$. As we will see, while simplistic, this model creates mechanically stable reference configurations in the infinitely persistent limit, and interesting glass-like (heterogeneous) dynamics in the presence of thermal noise.

\subsection{1D toy model}

To obtain estimates for appropriate simulation parameters and gain intuition for the expected steady-state behavior of our self propelled particle model in the limit of infinite persistence, we now consider a simple one-dimensional toy model. In a one-dimensional packing of $\tilde{N}$ monodisperse particles with radius $R$ and self-propulsion velocity $v_0$, take particle $i=0$ to be fixed at the origin. The other particles lie in the positive half of the number line and are governed by the equation of motion $\dot{x}_i = \frac{1}{\Gamma} \sum_j F_{i, \text{int}} - v_0$, where variables are defined similarly to above. The particles with $i>0$ have persistent velocities toward the origin. 

Given the condition for mechanical equilibrium in this toy (and our full) system, that the self-propulsion forces balance the repulsive interparticle forces, we can derive an expression for the typical pair overlap $(1 - \frac{r_{ij}}{R_{ij}})$ as a function of distance from the edge of the packing. Since this overlap is directly related to the local packing fraction, we choose simulation parameters such as the time step and $k$ to satisfy a constraint on the maximum packing fraction of the system, which occurs near the origin. Further, we predict the approximate form of an interaction pressure gradient which reveals the dependence of the static configurations on the simulation parameters $N$ and $v_0$. As will be discussed further below, this prediction for the interaction pressure also allows us to form a scaling relation for the stiffness associated with localized excitations in the interior of the active packings. The formulation of this toy model highlights that the distance $\chi$ from the free boundary is the most natural variable with which to examine structural gradients in the system. Considering the circular geometry of the packings generated by the full model (see Fig.~\ref{fig:intro}), we can approximate a radial slice of the 2D system using the 1D toy model. Further details regarding these calculations can be found in Appendix~\ref{sec:appB_toy}.

\subsection{Linear response and augmented Hessian framework}

Linear-response-based structural metrics are computed from curvatures of the potential energy landscape around a metastable minimum. For a material composed of $N$ interacting particles in $d$ dimensions, the total energy $U_{\text{int}}(\vec{X})$ is a function of the $Nd$-dimensional vector $\vec{X}$ representing points in coordinate space. Thus, the curvatures can be characterized by the Hessian, the matrix of second partial derivatives of the potential energy with respect to particle degrees of freedom: 
\begin{equation}
    \mathcal{M} = \frac{\partial^2U_{\text{int}}}{\partial \vec{X} \partial \vec{X}}.
\end{equation}
The dynamical matrix, used to compute the linear response, is computed strictly with respect to deformations from a stable reference configuration. Therefore, it is only well defined if such a stable reference configuration exists. However, when it is defined, the dynamical matrix is equivalent to the Hessian as derivatives with respect to particle positions and those with respect to deformations are identical. The eigenvectors and eigenvlaues of the Hessian constitute the spectrum of vibrational modes and associated stiffnesses of a solid if all particle masses are unity. Previous work on the mechanics of sheared athermal amorphous solids has demonstrated that a low-frequency population of these harmonic eigenmodes become quasi-localized under certain conditions, featuring a disordered core of large putative displacements on tens of particles decorated by a quadrupolar field which decays in magnitude as $r^{-(d-1)}$. These excitations are thus termed quasi-localized modes (QLMs), and identify glassy defects that become unstable under applied shear, generating structural rearrangements and non-affine motion \cite{lerner_micromechanics_2016,richard_universality_2020,richard_simple_2021}.  

For our initial ``point-of-interest" crowd simulations, the self-propulsion forces of the particles are infinitely persistent in the radial direction. Thus, there is an extra contribution to the total energy of the system which is exactly equivalent to a constant force spring potential pulling the particles toward the origin:
\begin{equation}
    U_{\text{ext}}(\vec{X}) = \Gamma v_0 \sum_i r_{i}, 
    \label{eq:extEnergy}
\end{equation}
where $r_{i}$ is the distance between particle $i$ and the origin. Therefore, we compute an ``augmented Hessian", where the energy has the usual contributions from interparticle interactions in addition to those from activity, which occur on the on-diagonal entries of the matrix: 
\begin{equation}
    \mathcal{M}_{\text{aug}} = \frac{\partial^2 (U_{\text{ext}}(\vec{X}) + U_{\text{int}}(\vec{X}))}{\partial \vec{X} \partial \vec{X}}.
\end{equation}
See Appendix~\ref{sec:appC_hessian} for details. In contrast to methods that probe the linear response and stability of active particle packings using approximations of the Hessian, our augmented Hessian framework is exact and requires only a snapshot of a static reference configuration.

\subsection{Static quantities}

Next, we describe two metrics for characterizing the static structure of the stable packings derived from the infinitely persistent limit of our self propelled particle model, interaction pressure and vibrability. Interaction pressure quantifies the distribution of forces in the active solid using a well-established Irving-Kirkwood description of the stress tensor \cite{yang_aggregation_2014, goldhirsch_microscopic_2002}. Vibrability is a linear-response-based structural metric that is used to quantify the propensity for local regions of the solid to deform under external deformation or active forcing \cite{tong_order_2014, tong_revealing_2019, richard_predicting_2020}.

The interaction pressure on particle $i$ is given by the trace $\sigma_{\text{int}}^{\alpha \alpha}$ of the interaction stress tensor whose components are a sum over the repulsive forces generated between $i$ and its neighbors:
\begin{equation}
    \sigma_{\text{int}}^{\alpha \beta} = \frac{1}{V_i} \sum_{\langle i j \rangle} F^{\alpha}_{ij} r_{ij}^{\beta},
    \label{eq:int_pressure}
\end{equation} 
where the sum is over (unique) neighbors of $i$, $F^{\alpha}_{ij}$ is the $\alpha$ component of the force of $j$ on $i$, $r^{\beta}_{ij}$ is the $\beta$ component of the distance vector pointing from $j$ to $i$, and $V_i$ is the volume associated with $i$ in a radical Voronoi tessellation of the system \cite{charbonneau_hopping_2014}. Since the Voronoi volumes of particles on the free boundary of the system are unbounded, they are excluded in the results that follow.

Vibrability was first defined in Ref.~\citenum{tong_order_2014} and uses the vibrational spectrum of the Hessian of a jammed packing to describe the susceptibility of particles to excitation and rearrangement. The vibrability of particle $i$ is given by
\begin{equation}
    \Psi_i = \sum_l^{dN-d} \frac{1}{\omega_l^2} \vert \vec{\psi}_{l,i} \vert^2,
    \label{eq:Psi}
\end{equation}
where the sum is over nonzero vibrational modes of the Hessian, $\omega_l$ is the frequency of mode $l$, and $\vert \vec{\psi}_{l,i} \vert^2$ is the squared magnitude of the polarization of particle $i$ in mode $l$. It was shown in Refs.~\citenum{tong_revealing_2019} and~\citenum{richard_predicting_2020} that vibrability is a good predictor of localized plastic rearrangements in sheared athermal computer glasses. In the augmented Hessian framework, we compute vibrability as in Eq.~\ref{eq:Psi}, but take the sum over the $dN-(d+1)$ nontrivial vibrational modes (as we discuss below) of the system. 

\subsection{Dynamic quantities}

We next explore the connection between the static structure of our active packings and their dynamics under small amounts of translational noise. In sheared amorphous solids at zero temperature, it is well-established that a population of microstructural defects 
are directly spatially correlated with future plastic deformation~\cite{richard_predicting_2020, richard_brittle--ductile_2021,wang_low-frequency_2019,patinet_connecting_2016,falk_dynamics_1998}. In contrast, in thermalized or active glasses it is generally difficult to demonstrate such a direct spatial correlation, except in non-molecular systems where the thermal fluctuations can be vanishingly small~\cite{chen_measurement_2011}. This is not unexpected; given a large population of underlying defects, various subsets of that population can be excited by thermal fluctuations or active forcing at any given time. Thus, resulting rearrangements of unstable regions occur sporadically, and so at any given time point regions with high mobility do not necessarily correlate strongly with structural indicator fields. To address this challenge, Schoenholz and collaborators~\cite{schoenholz_relationship_2017, schoenholz_structural_2016, sussman_disconnecting_2017, cubuk_unifying_2020, tah_quantifying_2021}  have developed a method that searches for structure-dynamics correlations by analyzing whether an indicator of structural softness defines a set of energy barriers that accurately predict the \emph{rate} of rearrangements.

We adopt this methodology here, analyzing particle rearrangement probabilities as a function of temperature $T$ and the structural indicator vibrability $\Psi$ (Eq.~\ref{eq:Psi}). As in previous work~\cite{schoenholz_relationship_2017, smessaert_distribution_2013,smessaert_structural_2014,candelier_spatiotemporal_2010}, we use a hop indicator to identify rearranging regions of the system. The indicator at a given time $t$ is computed directly from particle trajectory information with respect to two time intervals $A = \left[ t - t_R/2, t \right]$ and $B = \left[t, t + t_R/2 \right]$.  We take $t_R = 10$ in simulation time units, consistent with the work of Refs.~\citenum{smessaert_distribution_2013, schoenholz_structural_2016, sussman_disconnecting_2017} which chose $t_R \sim 10$ to correspond to the typical time taken for the system to complete a rearrangement in their simulations of Lennard-Jones polymer and bidisperse Kob-Andersen glasses. We have verified this choice independently by examining distributions of rearrangement times (determined as described below) in a representative range of simulations, where $t_R \sim 10$ constituted a reasonable upper bound for rearrangement time. Thus, $p_{i, \text{hop}}(t)$ for particle $i$ at time $t$ is given by:
\begin{equation}
    \label{eq:hop}
    p_{i, \text{hop}}(t) = \sqrt{ \langle \left( \vec{r}_i - \langle \vec{r}_i \rangle_B \right)^2 \rangle_A \: \langle \left( \vec{r}_i - \langle \vec{r}_i \rangle_A \right)^2\rangle_B},
\end{equation}
where $\langle \cdot \rangle_A$ and $\langle \cdot \rangle_B$ denote time averages over the specified intervals. We identify particle rearrangement events at the locations/times in which the hop indicator exceeds a threshold value, $p_{\text{thresh}} = 0.2$. Since we seek to identify rearrangements that result in irreversible structural changes, this threshold was chosen such that, for a representative set of example noisy simulations, the contact networks vary nontrivially for inherent structures computed with respect to configurations directly preceding and succeeding the times where $p_{\text{thresh}}$ is crossed.

In thermal systems, one expects that rearrangement rates are an Arrhenius function of energy barrier heights:
\begin{equation}
    P_R(S,T) = P_0(S)\exp\left( - \frac{E(S)}{T}\right),
\end{equation}
where $T$ is temperature, $P_0$ is a rearrangement attempt frequency, and $E$ is the energy barrier to rearrangement. Both $P_0$ and $E$ are generically functions of the structural indicator field $S$.

In Ref.~\citenum{schoenholz_relationship_2017}, $S$ is taken to be a machine-learning-derived softness field and the authors demonstrate that, after segmenting the system into bins of constant $S$, the dynamics are indeed Arrhenius with energy barriers that scale linearly with softness. In this work, we take $S$ to be the vibrability field, $\Psi$. Consistent with Ref.~\citenum{schoenholz_relationship_2017} and related works, in the discussion that follows we take an Arrhenius relationship between $P_{R}(\Psi)$ and $1/T$ in a given region to indicate that vibrability has successfully estimated the corresponding rearrangement energy barrier.

Additionally, since there are strong spatial gradients in hop indicator (as we demonstrate below) in our active packings, we compute $P_R(\Psi)$ by averaging particle rearrangement counts over multiple timesteps: 
\begin{equation}
    P_{R}(\Psi) = \frac{N_R(\Psi)}{N(\Psi)\cdot \Delta t_R},
\end{equation}
where $N_R$ is the number of rearranging particles with vibrability $\Psi$ in the time interval $\Delta t_R$, $N(\Psi)$ is the number of particles with vibrability $\Psi$, and $\Delta t_R$ is the time between rearrangements of particles with vibrability $\Psi$. $\ln(P_R(\Psi))$ measurements are computed and averaged for each selected value of $\Psi$ (computed from the appropriate static reference configuration) over the duration of each noisy simulation and over duplicate simulations with the same parameter ($N$, $v_0$, and $T$) choices.


\section{\label{sec:results}Results}

\subsection{Static packings}

\begin{figure}[hb]
\includegraphics[width = 0.97\linewidth]{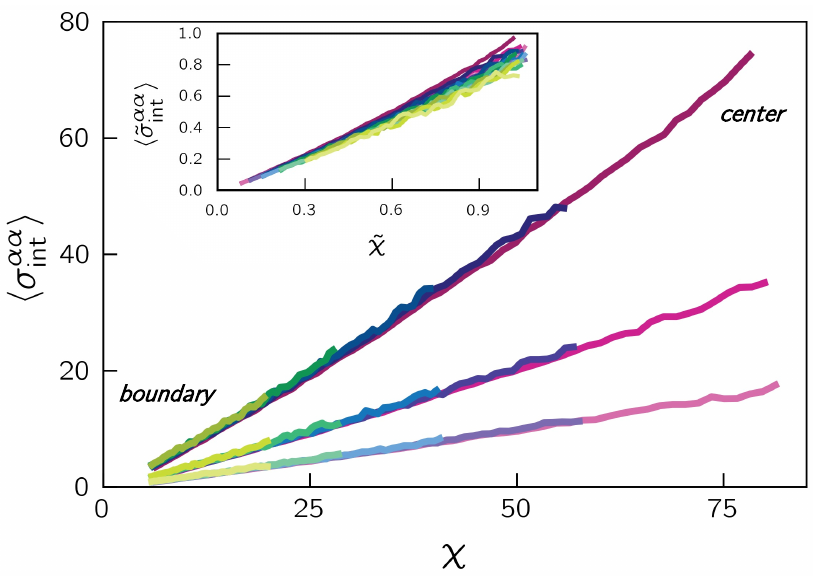}
\caption{\label{fig:pressure} Mean interaction pressure as a function of distance from the exterior of packings. Individual lines correspond to the mean pressure for 25 packing ensembles with $N \in \left\{256, 512, 1024, 2048, 4096 \right\}$ and $v_0 \in \left\{ 0.25, 0.5, 1.0 \right\}$. Each color represents a different system size $N$ (increasing from green (left) to magenta (right)) and each saturation level represents a different self propulsion velocity $v_0$ (increasing from light (bottom) to dark (top)). \textit{Inset}: Data rescaled according to the toy model discussed in the main text and Appendix~\ref{sec:appB_toy}, demonstrating an approximate collapse for all choices of simulation parameters $N$ and $v_0$.}
\end{figure}

To study the material properties and stability of our ensemble of reference configurations, we first examine structural features and gradients.  First, we computed the interaction pressure $\sigma^{\alpha\alpha}_{\text{int}}$ as a function of distance from the exterior of the packings. The 1D toy model introduced above and detailed in Appendix~\ref{sec:appB_toy} predicts that interaction pressure should increase monotonically with distance from the exterior of the system, $\chi$. The toy model further predicts that the scaling of interaction pressure with $ \frac{\chi}{\sqrt{N}\langle R \rangle}$ should be approximately independent of simulation parameter choice, $N$ and $v_0$, when rescaled by the quantity $\Gamma v_0 \sqrt{N} \langle R \rangle$ (where $\langle R \rangle \approx 1.2$ is the average particle radius and $\sqrt{N}\langle R \rangle$ is a good estimate for the radius of the packings). Thus, we define $\tilde{\chi} = \frac{\chi}{\sqrt{N} \langle R \rangle }$ and $\tilde{\sigma}^{\alpha \alpha}_{\text{int}} = \frac{\sigma^{\alpha \alpha}_{\text{int}}}{\Gamma v_0 \sqrt{N}\langle R \rangle}$. In Fig.~\ref{fig:pressure}, we show the mean interaction pressure as a function of $\chi$ as well as the rescaled mean interaction pressure $\tilde{\sigma}^{\alpha \alpha}_{\text{int}}$ as a function of $\tilde{\chi}$. As shown in the inset of the figure, these rescaled variables produce an approximate collapse of the data across the parameter range of our ensemble. The collapse is especially effective near the exterior (small $\tilde{\chi}$) -- for large $\tilde{\chi}$, there is noticeable deviation in the data that depends systematically on system size $N$. This feature is due to contributions to the interaction pressure $\sigma^{\alpha\alpha}_{\text{int}}$ by the Voronoi volume $V$ (see Eq.~\ref{eq:int_pressure}), which decreases monotonically with $\tilde{\chi}$. In our definition of the rescaled variables $\tilde{\sigma}^{\alpha\alpha}_{\text{int}}$ and $\tilde{\chi}$ motivated by the 1D toy model detailed in Appendix~\ref{sec:appB_toy}, these contributions from the Voronoi volume are neglected for simplicity.

\begin{figure}[ht]
\includegraphics[width = 0.70\linewidth]{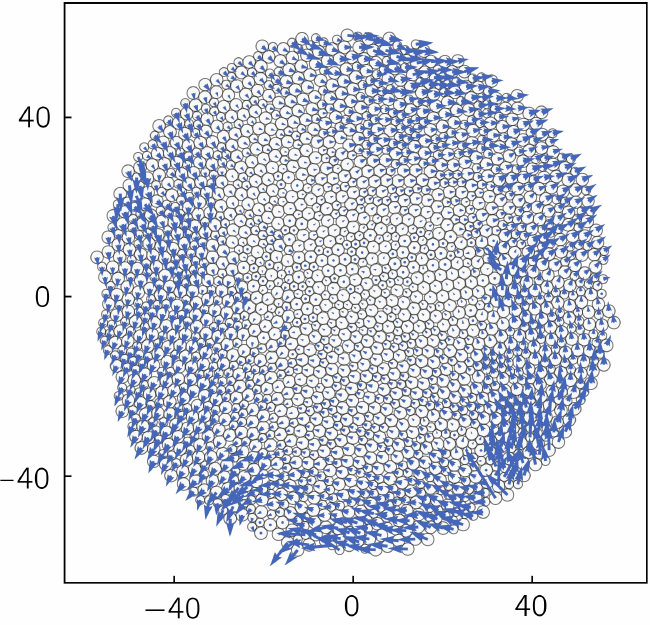}
\caption{\label{fig:mode} Sample low-frequency vibrational mode of the augmented Hessian for a static packing with $N=2048$ and $v_0=0.5$. The mode shows a large amount of collective motion around the exterior of the system compared to the interior.}
\end{figure}

\begin{figure*}[ht]
\includegraphics[width = 0.825\paperwidth]{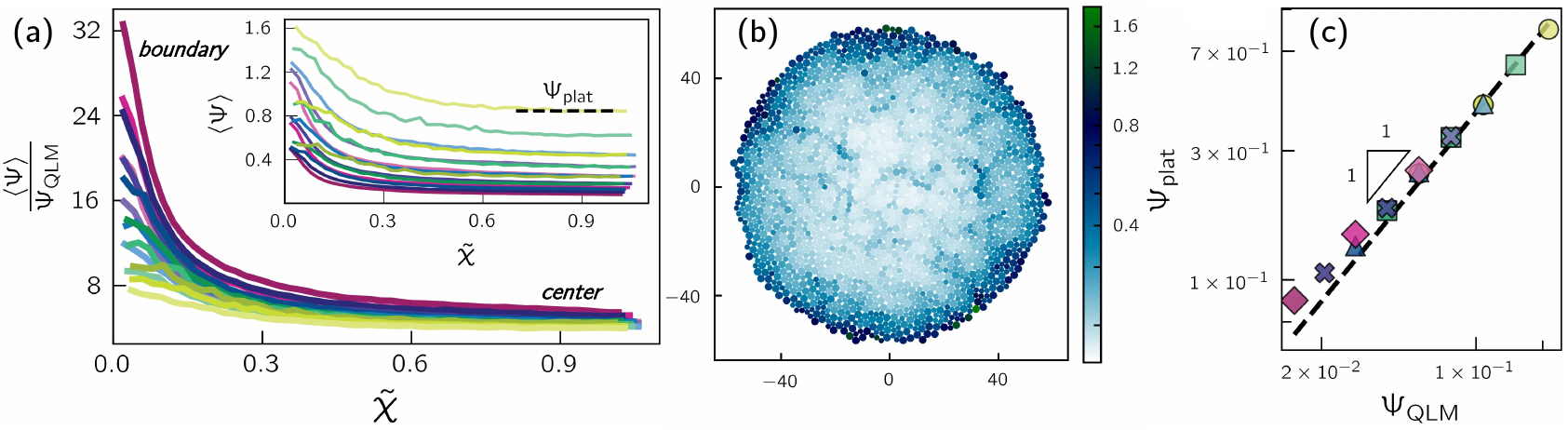}
\caption{\label{fig:vibrability} \textbf{(a)} Mean rescaled vibrability as a function of rescaled distance from the exterior of the packings, color scale as in Fig.~\ref{fig:pressure}. The inset shows the unscaled data and a sample value of $\Psi_{\text{plat}}$ as a guide to the eye. \textbf{(b)} The same configuration as in Fig.~\ref{fig:mode} with each particle colored by its vibrability. Particles on the interior are much less susceptible to rearrangements than those near the free boundary. \textbf{(c)} Measured vibrability plateau values $\Psi_{\text{plat}}$ compared to corresponding predicted plateau values $\Psi_{\text{QLM}}$. The dashed line of slope 1 indicates direct proportionality between $\Psi_{\text{QLM}}$ and $\Psi_{\text{plat}}$. Color scale as in (a), and each marker represents a different system size ranging from $N = 4096$ (diamonds, left) to $N=256$ (circles, right).}
\end{figure*}

Inspired by studies that utilize linear-response-related metrics to form structure-dynamics predictions as discussed above, we compute and diagonalize the augmented Hessian, examining the spectra of vibrational modes of our static reference configurations. We identify one rotational zero mode and two low-frequency trivial translational modes in the spectra of our packings. Typically, the vibrational spectra of solids have $d$ translational modes with zero frequency, but in our framework these modes have finite frequency due to the presence of the external potential (Eq.~\ref{eq:extEnergy}). In Fig.~\ref{fig:mode}, we show an example nontrivial low-frequency augmented Hessian eigenmode that is representative of typical soft modes for these systems. The mode exhibits wave-like motion emanating from the center of the packing as well as increased surface mobility. However, this vibrational mode does not show any characteristics of QLMs, as the putative collective displacement primarily involves a large number of particles on the exterior of the system and lacks the quadrupolar structure that has been shown to represent instabilities in traditional glassy systems \cite{richard_simple_2021,richard_universality_2020,lerner_micromechanics_2016}. Through direct examination, we have confirmed that such quasi-localized excitations are not commonly realized in the low-frequency modes of the augmented Hessians of our systems, especially near the interior of the packings. This is consistent with intuition we detail below regarding the stiffness of QLMs as a function of local pressure (see Appendix~\ref{sec:appD_barrier}). Further, it suggests that vibrability, a weighted sum over the soft modes of the augmented Hessian (Eq.~\ref{eq:Psi}), may struggle to predict rearrangement events on the high-pressure interiors of our packings.

The observation that excitations in the low-frequency regime of augmented Hessian spectra are concentrated near the edge of the system is highly reminiscent of previous work on jammed packings with free boundaries (Ref.~\citenum{msussman_disordered_2015}). Since the focus of our study is primarily to identify localized modes that predict plastic rearrangement in a specific class of modeled active solids, we do not develop such in-depth mode analysis here. However, we emphasize that our model differs significantly from previously analyzed systems, as our packings feature strong pressure gradients and those in Ref.~\citenum{msussman_disordered_2015} have homogeneous overall pressure. We explore the decay in vibrational magnitude exhibited in the low-frequency modes of the augmented Hessian in Appendix~\ref{sec:appE_mode}.

We next use our augmented Hessian spectra to develop structural indicator fields. Fig.~\ref{fig:vibrability}b shows a static configuration with $N = 2048$ and $v_0 = 0.5$ where each particle is colored by its vibrability, as defined by Eq~\ref{eq:Psi}. Similar to the low frequency vibrational modes themselves, the vibrability is large on the exterior and decreases quickly approaching the center of the packing. This trend is also depicted in the inset of Fig.~\ref{fig:vibrability}a where we show the mean vibrability as a function of $\tilde{\chi}$ for our ensemble of packings. 

Each vibrability profile reaches a plateau value at large $\tilde{\chi}$ in the interior. We hypothesize that this plateau value, $\Psi_{\text{plat}}$, is dominated by contributions from localized excitations whose vibrational frequencies depend on local pressure and thus on the simulation parameters $N$ and $v_0$. By combining the prediction for the pressure from the 1D toy model detailed in Appendix~\ref{sec:appB_toy} with a scaling relation for the stiffness of QLMs as a function of local pressure, we are able to generate a prediction for the vibrability of QLMs near the center ($\tilde{\chi}\sim 1$) of the packings, $\Psi_{\text{QLM}}$, as a function of $N$ and $v_0$. Details of this argument are discussed in Appendix~\ref{sec:appD_barrier}.

When the data are rescaled according to this prediction, as shown in the main panel of Fig.~\ref{fig:vibrability}a, the vibrabilty profiles exhibit an approximate collapse near the center of the packings ($\tilde{\chi} \sim 1$). On the exterior, there is more significant variation among the curves, which agrees with the interpretation that the vibrability in this regime has significant contributions from low-frequency, spatially decaying surface vibrations as depicted in Fig.~\ref{fig:mode}. 

Additionally, the large-$\tilde{\chi}$ collapse is poorer for systems with the largest local pressures (\textit{e.g.}~the dark magenta curve in Fig.~\ref{fig:vibrability}a corresponding to systems with $N=4096$ and $v_0 = 1.0$). To quantify the quality of this collapse as a function of $N$ and $v_0$, we plot the measured vibrability plateau values $\Psi_{\text{plat}}$ vs.~the predicted values $\Psi_{\text{QLM}}$ in Fig.~\ref{fig:vibrability}c (see Appendix~\ref{sec:appD_barrier} for details). As expected, this analysis highlights deviations of our scaling prediction from the actual vibrability plateau values $\Psi_{\text{plat}}$ at the largest values of $N$ and $v_0$. These deviations are likely due to increased multi-body interactions and larger local pressure fluctuations (that scale as $N^{1/2}$ and $v_0^1$ similarly to the pressure itself) which are not accounted for in our simple 1D model and vibrability scaling argument. In Appendix~\ref{sec:appF_overlaps}, we confirm that deviations from our scaling predictions are smaller in a system with higher particle stiffness $k$, consistent with this expectation.

\subsection{Dynamics in presence of translational noise}

\begin{figure}[ht]
\includegraphics[width = 0.97\linewidth]{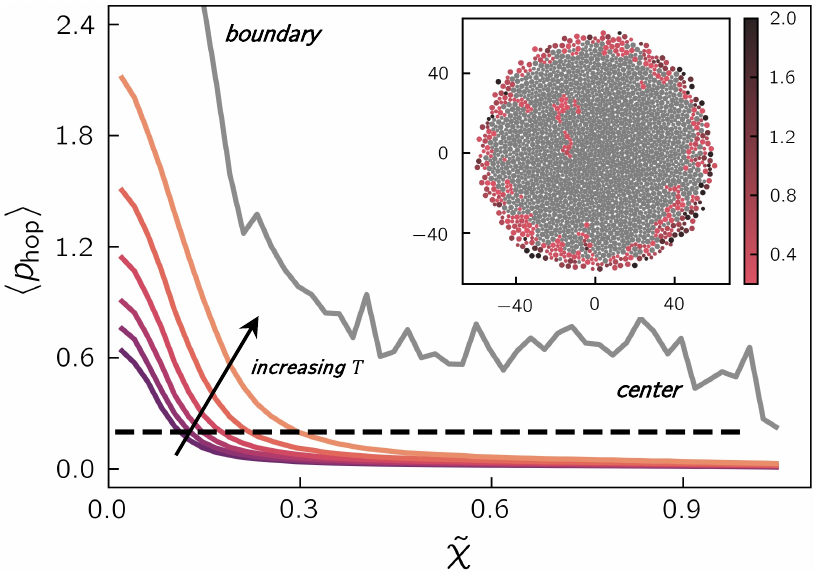} 
\caption{\label{fig:hop} Mean hop indicator as a function of rescaled distance from the exterior of the 25-duplicate ensemble of packings with $N = 2048$ and $v_0 = 0.5$. Each color represents a different temperature ranging from $T = $ 0.125 (bottom, dark) to 0.320 (top, light). The horizontal dashed line is placed at $p_{\text{thresh}} = 0.2$ to show the hop indicator threshhold which represents particle rearrangements. The grey line shows the maxiumum hop indicator over time and simulation duplicates for systems with $N = 2048$, $v_0 = 0.5$, and $T = 0.125 $. The inset shows an example configuration with $T = 0.197$ colored by threshholded hop indicator. Particles with $p_{\text{hop}} < p_{\text{thresh}}$ are colored grey and those with  $p_{\text{hop}} \geq p_{\text{thresh}}$ are colored according to the magnitude of the hop indicator. Notably, the maximum hop indicator profile and the snapshot in the inset show that rearrangements indeed occur throughout the entire depth of the packings.}
\end{figure}

Next, we examine the dynamics that result when thermal noise is added to the particle trajectories. Starting from the stable reference configurations discussed above, noise is added with magnitude controlled by the temperature $T \in \left\{ 0.125, 0.142, 0.165, 0.197, 0.244, 0.320 \right\}$. Using the hop indicator (Eq.~\ref{eq:hop} above) as a measure of particle mobility, we compare the rearrangement dynamics at different locations in the packings. As shown in Fig.~\ref{fig:hop}, for our ensemble of systems with $N=2048$ and $v_0 = 0.5$, there is a dramatic decrease in the mean hop indicator as a function of $\tilde{\chi}$ for all temperatures. These results are similar for other parameter ($N$ and $v_0$) choices. This dynamic profile is reminiscent of the vibrability profiles presented above, which predict increased mobility near the edge of the packings. 

Even though the majority of rearrangement events occur on or near the exterior of the packings, it is important to note that particles on the interior of the system do undergo occasional rearrangement. This can be seen from the snapshot in the inset of Fig.~\ref{fig:hop}, where particles are colored grey if $p_{i,\text{hop}} < p_{\text{thresh}}$, and colored according to the magnitude of the hop indicator if $p_{i,\text{hop}} \geq p_{\text{thresh}}$. Similarly, the grey curve in Fig.~\ref{fig:hop} shows the maxiumum of the hop indicator in different regions of the lowest temperature systems, which consistently exceeds $p_{\text{thresh}}$. 

\begin{figure}[h]
\includegraphics[width = 0.97\linewidth]{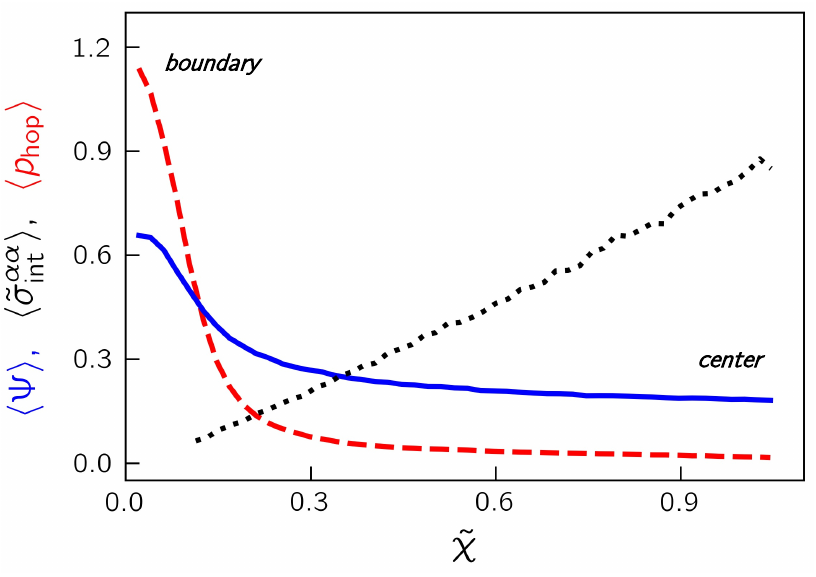}
\caption{ \label{fig:gradients} Structural and dynamic gradients as a function of rescaled distance from the exterior of the packings. Mean (rescaled) interaction pressure (dotted, black) and mean vibrability (solid, blue) are measured from the ensemble of static structures with $N=2048$ and $v_0 = 0.5$ and mean hop indicator (dashed, red) from the corresponding dynamics for $T = 0.197$.}
\end{figure}

The data we present here highlight important similarities to and differences from results regarding the structure and dynamics glassy thin films. Fig~\ref{fig:gradients} shows the average interaction pressure, vibrability, and hop indicator as a function of $\tilde{\chi}$, similarly to Fig.~1 in Ref.~\citenum{sussman_disconnecting_2017}. Notably, there are upticks in both hop indicator and vibrability near the free boundary, whereas the thin films studied in Ref.~\citenum{sussman_disconnecting_2017} exhibit only an analogous uptick in hop indicator. The thin film systems also exhibit little-to-no gradient in pressure. These differences highlight the utility of our augmented Hessian framework, and suggest that the pressure gradient in our active packings contributes significantly to their overall mechanical behavior.

\begin{figure}[h]
\includegraphics[width = 0.95\linewidth]{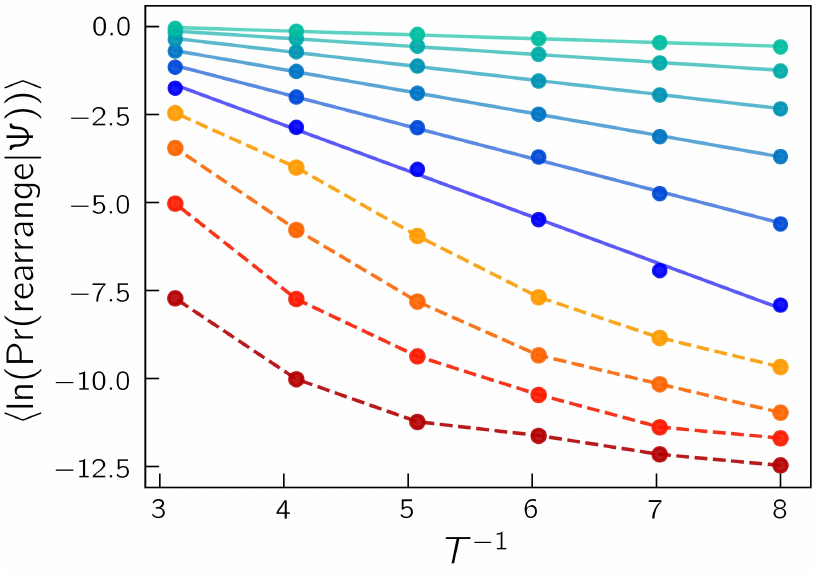}
\caption{\label{fig:arrhenius} Arrhenius plot for ensemble of packings with $N = 2048$, $v_0=0.5$, and the full range of temperatures we examined. The average log of the rearrangement probability in different bins of constant vibrability is plotted as a function of inverse temperature. There are 10 bins of vibrability ranging from $\Psi \approx 0.18$ (bottom) to $\Psi \approx 0.73$ (top). For vibrability bins where the relationship is well approximated by a linear fit (determined by an associated chi-squared value of less than $0.05$ ) (solid lines, cool colors), the rearrangement dynamics are Arrhenius, whereas bins exhibiting nonlinear trends represent sub-Arrhenius regions of the system (dashed lines, warm colors).}
\end{figure}

Lastly, we study rearrangement probabilities in our thermalized packings as a function of temperature and reference configuration vibrability $\Psi$ in order to determine whether rearrangement energy barriers are well-represented by $\Psi$. In Fig.~\ref{fig:arrhenius}, we show the mean of the natural log of the rearrangement probabilities in $n_\text{bin} = 10$ bins of approximately constant vibrability (independent of $\chi$) as a function of inverse temperature for our ensemble of systems with $N=2048$, $v_0 = 0.5$, and $T \in \left\{ 0.125, 0.142, 0.165, 0.197, 0.244, 0.320 \right\}$. Strikingly, the behavior is Arrhenius for large values of $\Psi \gtrsim 0.22$, but sub-Arrhenius for small values of $\Psi \lesssim 0.22$. This result is qualitatively independent of $N$ and $v_0$ and varies significantly from the results of Refs.~\citenum{schoenholz_structural_2016},~\citenum{sussman_disconnecting_2017}, and~\citenum{tah_quantifying_2021}, which identified Arrhenius behavior for individual values of softness in both bulk, thin film, and active/biological systems. Recalling the static structural gradients (vibrability and interaction pressure) above, we notice that the sub-Arrhenius portions of the system lie in the interior of the packing, where the interaction pressure is high and soft modes are suppressed. 

Taken together, our results indicate that dynamic rearrangements in the interior of packings are generally \textit{not} well-predicted by our augmented Hessian framework and that the vibrability alone is not a good structural indicator in the interior of this system. We note that useful information may still exist, for instance, in local variations in vibrability (which we preliminarily examined and found that it did not correlate well with rearrangement events), but our work suggests that other approaches such as non-linear-response-based metrics will be more fruitful, as discussed below.


\section{\label{sec:discussion} Discussion and Conclusions}

In this work, we studied results from computer simulations of a soft active particle model in two dimensions with directed self-propulsion in the overdamped regime. We analyzed static structures that are formed in the infinitely persistent limit of the activity and found a strong pressure gradient that is consistent with a simple 1D toy model. We then developed an augmented Hessian to capture the active forces in our analysis of the vibrational properties of the system. Further, we used these eigenspectra and a structural indicator, vibrability, to estimate rearrangement energy barriers in analogy to previous work on the dynamics of supercooled liquids and sheared amorphous solids. Then, we observed the resulting dynamics when simulations with translational noise are initiated from the static configurations. Similar to other particle-based systems with free boundaries, we measured a gradient in mobility that persists through the depth of the packings and features enhanced mobility near the free boundaries. We found that, near the boundary of the packings, vibrability is a good structural indicator of energy barrier heights and rearrangement probabilities, but it fails to represent these features in the interior of the systems.

This failure is surprising, as previous work by Bottinelli and collaborators \cite{bottinelli_can_2018,bottinelli_emergent_2016,bottinelli_how_2017,bottinelli_when_2019} suggested that vibrational analyses of real crowds and crowd models, estimated from a matrix of time-averaged two-particle correlation functions, were able to forecast localized rearrangements and wave-like motion. Similary to the results which we present here, in Ref.~\citenum{bottinelli_emergent_2016}, the authors note that simulated half-circular point-of-interest crowds exhibit approximately linearly increasing pressure approaching a point of interest. However, by examining a small number of low-frequency vibrational modes ($\sim 3 \%$ of the spectrum) derived from an approximation of the Hessian, Bottinelli \textit{et.~al.}~identified localized soft regions on the interior of their packings which directly spatially correlated with increased noise-induced particle mobility. Distinctly, our results suggest that such correlations are quite difficult to draw from linear-response-based analyses in regions of high local pressure.

Our analysis suggests an updated interpretation of the results of Refs.~\citenum{bottinelli_emergent_2016} and~\citenum{bottinelli_can_2018}. Recently, it has been shown in both theoretical works and experiments that the dynamics of active systems do not obey the fluctuation-dissipation theorem (FDT) \textit{a priori}~\cite{mizuno_nonequilibrium_2007, lau_microrheology_2003, burkholder_fluctuation-dissipation_2019, fodor_how_2016}. Furthermore, to construct the correlation functions necessary to approximate the Hessian in the framework of Bottinelli and coworkers, one must indeed time-average over dynamics where the contact network underlying the system has changed; thus, averages are taken over multiple metastable states. As we outlined in Sec.~\ref{sec:intro}, work by Henkes \textit{et.~al.}~\cite{henkes_extracting_2012} suggests that this approximation method only holds under specific conditions regarding the dynamics of a system (that they satisfy FDT) and the existence of a well-defined, time-invariant solid reference state underlying its structure. When these conditions are not met, the correspondence of the approximation to the real linear response of the system might actually be quite poor. 

Further, our study demonstrates that the real (augmented) Hessian cannot predict rearrangements in high pressure regions. Therefore, we speculate that the time-averaged approximation picks up dynamic features that are not present in the exact Hessian itself, and that these features are important for the predictive capability of the method of Bottinelli \textit{et.~al.}. Future work might further compare these approaches to enhance our overall forecasting capability surrounding the structure and dynamics of complex active solids. In fact, the model examined in our study would be appropriate for a direct comparison between these approximate and exact approaches. 

Additionally, we note that even though the choice of the simple soft sphere model described above (and in Ref.~\citenum{bottinelli_emergent_2016}) is rather artificial and may not accurately represent the interactions in real human crowds, these studies serve as important initial explorations of the connection between structure and dynamics in active solids. If our framework is to be used to understand and control the behavior of real crowds, more consideration should be given to determining ``effective potentials" that might govern pairwise interactions between human beings as well as interactions between humans and external stimuli (such walls, points of interest, and other environmental factors) \cite{kim_velocity-based_2015,warren_collective_2018}.

Although our work here focused on ``point-of-interest" active crowds, since the directed activity can be mapped onto an external potential, our observations are likely relevant to other classes of systems with pressure gradients and self-generated boundaries. For example, particle aggregates formed under microgravity conditions exhibit a spherical profile, gradients in density, and a free boundary \cite{love_particle_2014,kothe_free_2013}. Similarly, recent studies investigating the relaxation of active colloidal glasses attained sedimentation by inclining the experimental set-up at a small angle. As a result of this geometry, Klongvessa \textit{et.~al.}~measured a gradient in density at all levels of particle activity and resultant gradients in mobility \cite{klongvessa_active_2019,klongvessa_nonmonotonic_2019} Additionally, a number of works study granular shear flow and shear banding in cylindrical Couette-like geometries under varying gravitational strength and/or confining pressure \cite{murdoch_granular_2013,voth_ordered_2002}. Experimental and simulated systems with this set-up exhibit localized particle rearrangements in the presence of density heterogeneity. Further, experiments on particulate systems that are driven by external magnetic fields or vibrations exhibit enhanced surface mobility and glassy dynamics characterized by the coexistence of populations of particles with arrested dynamics and those that undergo large displacements via occasional neighbor exchanges (referred to as ``dynamical heterogeneity" in some literature) \cite{voth_ordered_2002, thomas_structures_2004, escobar_glass-_2020}. Lastly, we note that the structure and dynamics of sand and grain piles are largely dominated by the presence of pressure gradients \cite{roul_simulation_2010, ai_finite_2013}.

A relevant question to our discussion is whether facilitated dynamics occur in systems with free boundaries, where enhanced mobility and frequent structural changes on the exterior of the packings could facilitate other nearby rearrangements that propagate in toward the center over time. In Ref.~\citenum{sussman_disconnecting_2017}, Sussman \textit{et.~al.}~measure a softness propagator which suggests that facilitation does not explain enhanced surface mobility in glassy thin films. A similar analysis might be interesting in systems that also have strong pressure gradients.

In general, our results suggest that an augmented Hessian framework could be directly applied to forming structure-dynamics predictions in any solid-like system for which one can i) define suitable a reference configuration and ii) write down a twice-differentiable augmented potential energy that completely captures the characteristics of any internally-generated active forces or external applied fields. Still, as we have shown, there are material-dependent subtleties that effect the predictive power of our technique such as the influence of global pressure gradients and boundary conditions on material stability. While most analyses involving linear-response-based structural metrics have been applied to mechanically stable systems where the Hessian is positive-definite, recent work investigating avalanche dynamics in sheared amorphous solids suggests that Hessians describing unstable systems may also be useful for predicting dynamics \cite{stanifer_avalanche_2021}. 

Last, we note that our methods are likely applicable to systems with different types of noise. While we focused here on a system with translational (additive) noise  -- the stochastic term $\eta_i$ in the equation for particle positions, Eq.~\ref{eq:noisy_eom} -- many studies of active matter focuses on systems with rotational (multiplicative) noise, where a stochastic term is instead added to the angular dynamics, Eq.~\ref{eq:angle_eom}. In the absence of interactions, such dynamics generate particles that execute persistent random walks. We have performed some preliminary simulations indicating that the dynamics in dense active crowd simulations with rotational noise are remarkably similar to those presented here for systems translational noise, especially when the noise magnitude is not too large. This suggests that our methods may be used to analyze systems with finite persistence times, which provides an interesting avenue for future work.

Given our observation that linear-response-based structural indicators fail in systems with strong pressure gradients, an obvious next question is how to formulate a better-performing predictive framework. In a number of recent works, a class of novel non-linear-response-based structural indicators have been constructed that address many of the shortcomings of simple linear-response-based metrics \cite{lerner_micromechanics_2016,kapteijns_nonlinear_2020,gartner_nonlinear_2016-1,gartner_nonlinear_2016,richard_simple_2021,richard_predicting_2020}. Namely, these so-called nonlinear plastic modes (NPMs) and their approximations have been shown to be robust representations of QLMs, the microstructural entities that control rearrangements in disordered solids. Importantly, these methods can quantify the asymmetry of the energy landscape. Thus, even if a mode has very high curvature -- so that the mode does not appear in the low-frequency harmonic spectrum --  it can still have a low energy barrier provided the mode is highly asymmetric.

Therefore, NPMs are a very promising future avenue for constructing a non-linear-response-based structural metric that successfully predicts rearrangements in systems with gradients in interaction pressure. As we detail in Appendix~\ref{sec:appD_barrier}, a simple scaling argument can be constructed which suggests that the stiffness of rearrangement-inducing excitations (QLMs), increases quickly with local pressure. This provides a potential explanation as to why vibrability derived from the augmented Hessian is not sensitive to QLMs that exist in the interior of our active packings, and highlights why NPMs are promising. Alternate methods for computing structural indicators could include machine learning approaches, where it will be important to determine how best to handle the strong gradients in pressure during the supervised learning phase. Overall, our work has elucidated that structural indicators for systems with pressure gradients should not be based on linear response alone.


\section*{Author Contributions}
All authors contributed to the overall conceptualization of this work. J.A.G. and E.M.S. wrote and developed the simulation and analysis code. J.A.G. ran the simulations and prepared the figures. J.A.G. and M.L.M. wrote the original draft of the manuscript. All authors contributed to reviewing and editing the manuscript. 

\section*{Conflicts of interest}
There are no conflicts to declare.

\section*{Acknowledgements}
We acknowledge a fruitful conversation with David Richard suggesting a scaling relationship between quasilocalized modes and the vibrability. We thank David Richard and Daniel Sussman for their useful comments on the manuscript. J.A.G. and M.L.M. acknowledge support from the Simons Foundation grant \#454947 and National Science Foundation NSF-DMR-1951921.  E.M.S. acknowledges support from MURI N00014-20-1-2479.


\appendix
\section*{Appendices}

\section{Model details}\label{sec:appA_model}

The equations of motion described above for the dynamics of our self propelled particle model in the limit of infinite persistence and in the presence of translational noise (Eqs.~\ref{eq:static_eom} and~\ref{eq:noisy_eom}) were integrated using the velocity Verlet method. To ensure numerical stability, during the formation of static reference configurations (in the absence of noise), we used a variable timestep proportional to the maximum unbalanced force in the system. These static simulations were run until the maximum unbalanced force reached a threshold of $\vert \vec{F}_{\text{unbalanced}} \vert < 10^{-8}$. In the noisy simulations, random numbers were drawn from a Gaussian with zero mean and unit variance, and rescaled by the variance of the noise, proportional to $\sigma \sqrt{dt}$ where $dt = 10^{-3}$ is the simulation time step, which was held constant \cite{volpe_simulation_2013,callegari_numerical_2019}. See the description of the 1D toy model below for a description of how this time step was chosen. In both static and dynamic simulations, particles whose positions were very close (within the precision of the simulation) to the central point of attraction were pinned to that location to prevent trivial fluctuating dynamics and numerical instability in the case of the static simulations. The simulation was implemented in Python, and just-in-time compiled with Numba \cite{noauthor_numba_nodate} to increase performance. 

\section{1D toy model}\label{sec:appB_toy}

In this appendix, we closely examine the one dimensional toy model mentioned in the main text which we use to pick appropriate simulation parameters and make general predictions about the structural features of our static packings. The model consists of a one-dimensional packing of $\tilde{N}$ monodisperse particles with radius $R$ and self-propulsion velocity $v_0$. The $0$th particle is fixed at the origin, and the other particles lie in the positive half of the number line and are governed by the equation of motion $\dot{x}_i = \frac{1}{\Gamma} \sum_j F_{i, j}^{\text{int}} - v_0$. Similarly to our full model, the force between two overlapping particles, $F_{i,j}^{\text{int}}$, is given by $F_{i, j}^{\text{int}} = - \frac{\partial \phi^{\text{int}}}{\partial x_i}$ with $\phi^{\text{int}}(x_{ij}) = \frac{k}{\alpha}(1 - \frac{x_{ij}}{2R})^{\alpha}$ where $x_{ij} = \vert x_j - x_i \vert$ is the distance between the particle centers and $\alpha = 2.5$ for Hertzian soft spheres. Using the condition for force balance in the system, that the interparticle forces must cancel the (cumulative) self-propulsion forces, we obtain an expression for the force $F_{i-1,i}^{\text{int}}$ between two adjacent particles: 
\begin{equation} \label{eq:toy_force}
    F_{i-1,i}^{\text{int}} = \Gamma v_0 (\tilde{N}-i).
\end{equation}
This expression also highlights that $\chi$, the distance from the exterior of the packing to a particle's center, is a natural variable in which to express structural and dynamic gradients of the system. 

We will first estimate the maximum overlap in this toy system to identify an appropriate choice for the simulation parameter $k$. By setting $i = 1$ and taking $\tilde{N}$ to be large, we can approximate the maximum overlap $\gamma_{i,j} = 1 - \frac{x_{ij}}{2R}$ in the 1D packing, as $F_{0,1}^{\text{int}} = \frac{k}{2R}(\gamma_{0,1})^{\alpha -1} \approx \Gamma v_0 \tilde{N}$ implies that 
\begin{equation}\label{eq:toy_overlap}
   \gamma_{0,1} \approx \left(\frac{2R\Gamma v_0 \tilde{N}}{k}\right)^{\frac{1}{\alpha - 1}}. 
\end{equation}
For a packing of crystalline monodisperse spheres of radius $R$ in 2D, the packing fraction $\phi$ can be estimated for $\gamma < 0.5$ via $\phi \approx \frac{\pi}{2\sqrt{3}}\frac{1}{(1-\gamma)^2}$. Applying the estimate for $\gamma_{0,1}$ to a circular packing in 2D, we first assume that radius of such a packing is approximately $2\tilde{N}R$. $\tilde{N}$ is again the number of particles in an analogous 1D packing representing a radial slice of the 2D system. Thus, the area of the 2D packing is $A\approx 4\pi\tilde{N}^2 R^2$. We can also approximate the 2D packing area by $A\approx N \pi R^2$ where $N$ is the total number of particles in the system. Equating these, we obtain $\tilde{N} \approx \frac{\sqrt{N}}{2}$ and $\gamma_{0,1} \approx (\frac{\Gamma R v_0 \sqrt{N}}{k})^{\frac{1}{\alpha -1}}$. Using this relation for the overlap and the above approximation of the packing fraction, we choose $k$ such that the maximum packing fraction in the center of the largest system, $N=4096$, does not exceed $\sim 1.3$, yielding $k \sim 1500$. We note here that the above arguments suggest that $\sqrt{N}R$ is a good approximation for the overall radius of our 2D packings, which motivates the definitions of the rescaled variables $\tilde{\chi}$, $\tilde{\Pi}$, and $\tilde{\sigma}^{\alpha\alpha}_{\text{int}}$. 
\begin{figure}[ht]
\includegraphics[width = 0.95\linewidth]{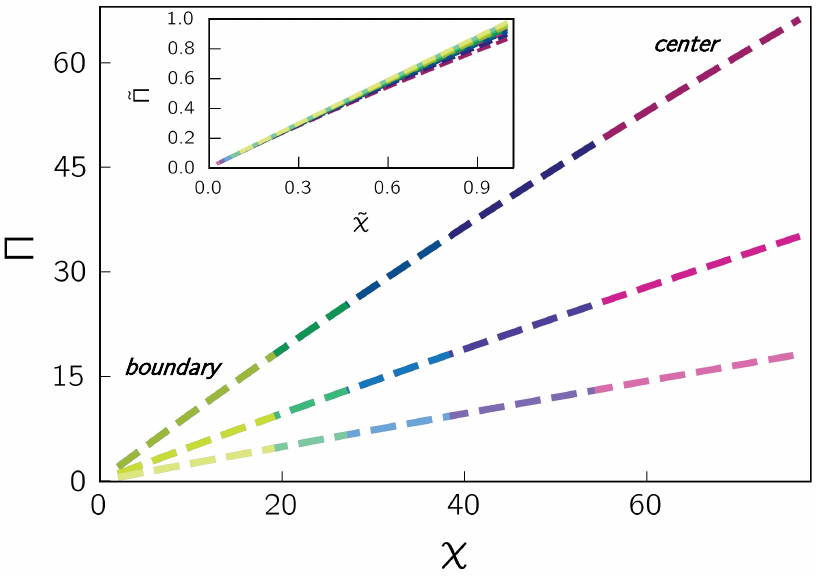}
\caption{\label{fig:append_toy} Toy-model analog to Fig.~\ref{fig:append_pi} below. $\Pi$, the approximated interaction pressure, is shown as a function of $\chi$. In the inset, $\Pi$ and $\chi$ are rescaled to $\tilde{\Pi} = \frac{\Pi}{\Gamma v_0 \sqrt{N} \langle R \rangle}$ and $\tilde{\chi} = \frac{\chi}{\sqrt{N} \langle R \rangle}$ according to Eq.~\ref{eq:toy_pi}, showing a collapse close to the free boundary of the system. The color map is the same as that of Fig.~\ref{fig:pressure} in the main text.}
\end{figure}

\begin{figure}[h]
\includegraphics[width = 0.95\linewidth]{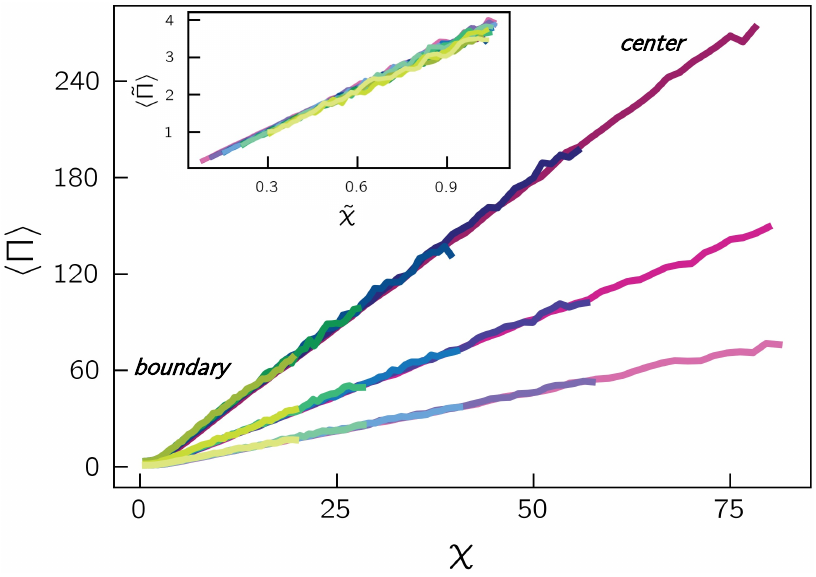}
\caption{\label{fig:append_pi} Simulation data corresponding to the toy-model prediction depicted in Fig.~\ref{fig:append_toy} above. Color map as in Fig.~\ref{fig:pressure} in the main text. $\Pi$, the approximated interaction pressure, is shown as a function of $\chi$. The inset shows the rescaled approximate pressure $\tilde{\Pi}$ as a function of $\tilde{\chi}$.}
\end{figure}

Next, using the above expressions, we can compute $\Pi_i = F_{i-1,i}\:x_{i,i-1}$ as a function of $i$, which corresponds approximately to individual contributions to interaction pressure as a function of distance from the exterior of the packing in this toy model. Using Eqs.~\ref{eq:toy_force} and~\ref{eq:toy_overlap} (slightly modified to express the overlap as a general function of $i$), we have:
\begin{multline}
    \Pi_i = F_{i-1,i}\:x_{i-1,i} = \\ 2R\Gamma v_0 (\tilde{N}-i)\left(1 - \left(\frac{2R\Gamma v_0 (\tilde{N}-i)}{k} \right)^{\frac{1}{\alpha-1}} \right).
\end{multline}
Considering simple geometric arguments to transform this into a function of $\chi$, we finally obtain:
\begin{equation}\label{eq:toy_pi}
    \Pi(\chi) = \Gamma v_0 \chi  \left(1 - \left(\frac{\Gamma v_0 \chi}{k} \right)^{\frac{1}{\alpha-1}} \right).
\end{equation}
This function is plotted for a realistic range in $\chi$ and for appropriate parameter choices in Fig.~\ref{fig:append_toy}. For direct comparison to our simulation data, Fig.~\ref{fig:append_pi} shows $\Pi$ ($\tilde{\Pi}$) as a function of $\chi$ ($\tilde{\chi}$) for the same ensemble of simulations as Fig.~\ref{fig:pressure} in the main text. Clearly, the toy model (including the relevant rescaled variables $\tilde{\chi}$, $\tilde{\Pi}$, and $\tilde{\sigma}^{\alpha\alpha}_{\text{int}}$) succeeds in capturing the behavior of the static packings produced in our full 2D model.

\begin{figure}[ht]
\includegraphics[width = 0.95\linewidth]{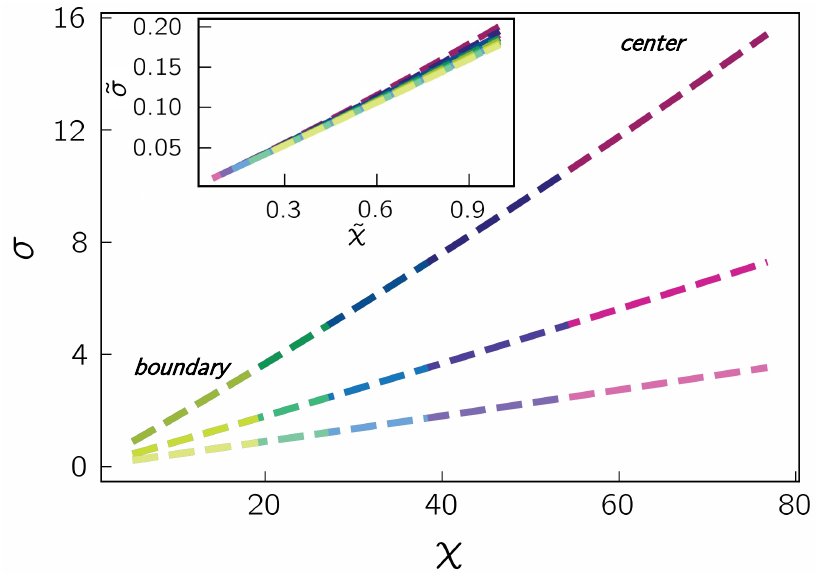}
\caption{\label{fig:append_sigma} Toy-model analog to Fig.~\ref{fig:pressure} in the main text. $\sigma$, the interaction pressure, is shown as a function of $\chi$. In the inset, $\sigma$ and $\chi$ are rescaled to $\tilde{\sigma} = \frac{\sigma}{\Gamma v_0 \sqrt{N} \langle R \rangle}$ and $\tilde{\chi} = \frac{\chi}{\sqrt{N} \langle R \rangle}$ according to Eq.~\ref{eq:toy_pi}, showing a collapse close to the free boundary of the system.}
\end{figure}

The definition of $\Pi$ in Eq.~\ref{eq:toy_pi} above differs from Eq.~\ref{eq:int_pressure} for $\sigma^{\alpha\alpha}_{\text{int}}$ by a factor of the Voronoi volume $V$ associated with a given particle. Since $\sigma^{\alpha\alpha}_{\text{int}}$ is an intensive variable commonly examined in literature studying jammed packings and active systems, we focus on it primarily in the main text, despite the simplicity of $\Pi$ in our toy model. Thus, here, we estimate particle Voronoi volume in the framework of our 1D toy model by examining typical interparticle distances $x_{i-1,i}$. We obtain: 
\begin{equation}\label{eq:toy_V}
    V(\chi) = \langle R \rangle \left[ 2+ \left( \frac{\Gamma v_0 \chi}{k}\right)^{\frac{1}{\alpha - 1}}\left( -1 - \left(1-\frac{2 \langle R \rangle }{\chi} \right)^{\frac{1}{\alpha-1}} \right)\right].
\end{equation}
The full estimate for the interaction pressure $\sigma$ in our 1D toy model is thus given by the quotient of Eqs.~\ref{eq:toy_pi} and~\ref{eq:toy_V}. This expression for $V$ suggests that in our systems, the Voronoi volume associated with a particle decreases monotonically with $\tilde{\chi}$ for all $N$ and $v_0$. Further, larger packings achieve smaller overall values of $V$ (due to increased particle density) as $\tilde{\chi} \rightarrow 1$. Thus, $\sigma^{\alpha\alpha}_{\text{int}}$ grows with $N$ near $\tilde{\chi} = 1$ faster than $\Pi$ does. This trend can be seen in Fig.~\ref{fig:append_sigma}, where we show $\sigma(\chi)$ and $\tilde{\sigma}(\tilde{\chi})$ computed in our toy model, in direct comparison to Fig. \ref{fig:pressure} in the main text.

Last, we choose a stable simulation time step by considering the maximum force generated between two particles in one timestep. In our toy model of monodisperse spheres in one dimension, if the $0$th particle and $1$st particle satisfy $x_{01}=2R$ at time $t$, the largest amount of overlap that can be generated via particle $1$'s self propulsion at time $t+dt$ is given by $\frac{v_0 dt}{2R}$. Thus, if we demand that the corresponding force generated by this overlap be less than some multiple $\epsilon$ of the self-propulsion force, we obtain an inequality for the simulation time step, $dt < (\frac{\epsilon v_0 \gamma 2 R}{k})^{1/(\alpha - 1)}(\frac{2R}{v_0})$. For our choice of simulation parameters and $\epsilon \sim 1\%$, this gives a timestep of $dt \sim 10^{-3}$. 

\section{Augmented Hessian} \label{sec:appC_hessian}

As discussed in the main text, a key result of our work is the formulation of the augmented Hessian framework. By exactly mapping directed self propulsion in our static packings to an external potential, we account for the contributions of active forces to the energy of the system. We compute $\mathcal{M}_{\text{aug}}$ from this total potential energy, and examine the corresponding soft modes. In this appendix, we compute the augmented Hessian for a general external potential. 

Consider the total potential energy of the system, given by $U(\vec{X}) = U_{\text{int}}(\vec{X}) + U_{\text{ext}}(\vec{X})$ as described above. Taking the second derivative of $U$ with respect to two degrees of freedom $x_{i\alpha}$ and $x_{j\beta}$ (with Latin indices corresponding to particles and Greek indices corresponding to spatial coordinates), we obtain a general expression for an element of the augmented Hessian. 

\begin{multline}
    \mathcal{M}_{\text{aug}, ij\alpha\beta} = \frac{\partial^2 U}{\partial x_{i\alpha} \partial x_{j\beta}} = \\ \sum_{\langle ij \rangle} \left[ \frac{\partial \phi_{\text{int}}}{\partial{r_{ij}}} \frac{\partial^2r_{ij}}{\partial x_{i\alpha} \partial x_{j\beta}} + \frac{\partial^2 \phi_{\text{int}}}{\partial^2 r_{ij}} \frac{\partial r_{ij}}{\partial x_{i\alpha}} \frac{\partial r_{ij}}{\partial x_{j\beta}} \right] + \\ \sum_{k,\gamma,\lambda} \left[ \frac{\partial^2 \phi_{\text{ext}}}{\partial x_{k\gamma} \partial x_{k\lambda}}\delta_{ki}\delta_{\gamma \alpha} \delta_{kj} \delta_{\gamma\beta} \right],
\end{multline}
where $U_{\text{int}} = \sum_{\langle ij \rangle} \phi_{\text{int}}(r_{ij})$ is a sum over energies of interacting pairs and $U_{\text{ext}} = \sum_k \phi_{\text{ext}}(x_{k\gamma})$ is a sum over external potential energies of individual particles. Given the form of the second term of this equation, it is clear that the external potential only has nonzero contributions to the augmented Hessian on the block diagonal terms of the matrix. 

\section{QLM stiffness scaling relation}\label{sec:appD_barrier}
To provide intuition for the utility of NPMs in future work that seeks to identify localized instabilities in disordered and active packings with unique structural features such as pressure gradients and free boundaries, we formulate a scaling relation for the stiffnesses associated with quasi-localised excitations (QLMs) in systems with varying homogeneous pressure based off of a body of work that studies the micromechanics of computer glasses. In Ref.~\citenum{gartner_nonlinear_2016}, Gartner \textit{et.~al.}~define $\kappa_{\vec{z}} \equiv \mathcal{M} : \vec{z} \vec{z} \sim \omega_{\vec{z}}^2$, the stiffness associated with the mode $\vec{z}$. Next, in Refs.~\citenum{lerner_characteristic_2018} and~\citenum{rainone_pinching_2020}, the authors examined local deformations in model glasses and identified a characteristic energy scale associated with quasi-localized excitations, which can be given by $\omega_{\text{QLM}} \sim \omega_g \sim \frac{c_s}{\xi_g}$ where $\xi_g$ is a glassy length scale and $c_s$ is the shear wave speed that scales with the overall pressure $p$ of jammed packings as $c_s\sim p^{1/4}$. Last, through examining sample-to-sample fluctuations in the shear moduli of computer glasses with short-range attractive potentials, González-López \textit{et.~al.}~showed in Ref.~\citenum{gonzalez-lopez_mechanical_2021} that the length scale $\xi_g$ changes with pressure as $\xi_g \sim p^{-1/2d}$ where $d$ is the number of spatial dimensions of the glass. 

Combining the above, we finally obtain a scaling prediction for $\omega_{\text{QLM}}$ with pressure, $\omega_{\text{QLM}} \sim p^{(d+2)/4d}$. Thus, $\kappa_{\text{QLM}} \sim p$ for QLMs in 2D systems with homogeneous pressure. Taken in the context of our results above which suggest that this type of localized instability is difficult to identify in the harmonic approximation for systems with strong pressure gradients, we expect that rearrangements could be more effectively predicted in future studies by searching for modes with very high asymmetry in the potential energy landscape that may be quite stiff relative to typical soft modes. In passing, we note that numerical studies investigating the vibrational modes of glasses approaching the unjamming transition have predicted that the density of QLMs decreases sharply with increasing overall pressure \cite{shimada_spatial_2018}.

Now, we use this scaling relation for $\omega_{\text{QLM}}$ to formulate the prediction for the plateau vibrabilities shown in Fig.~\ref{fig:vibrability}. Recall Eq.~\ref{eq:Psi} above for the vibrability. Given the spatial features of QLMs, namely that they feature large polarization vectors on a small number of particles, we conclude that the squared polarization magnitudes in the sum for vibrability are of order one for QLMs. Thus, contributions to vibrability by QLMs are dominated by their inverse squared frequencies. Since we showed above that $\omega_{\text{QLM}} \sim p^{1/2}$ in 2D, we conclude that $\Psi_{\text{QLM}} \sim p^{-1}$. Using Eq.~\ref{eq:toy_pi} as an estimation of the local pressure, we compute $\Psi_{\text{QLM}}$ for the appropriate parameters in our model when $\tilde{\chi} \sim 1$. Since low-frequency, wave-like vibrational modes decay quickly in magnitude away from the exterior of our packings (see Fig.~\ref{fig:mode} and Appendix~\ref{sec:appE_mode} below), it is likely that $\Psi_{\text{QLM}}$ is the most dominant contribution to vibrability in the interior. Note that this prediction is thus not valid for small $\tilde{\chi}$. 

\section{Mode analysis}\label{sec:appE_mode}
\begin{figure}[ht]
\includegraphics[width = 0.95\linewidth]{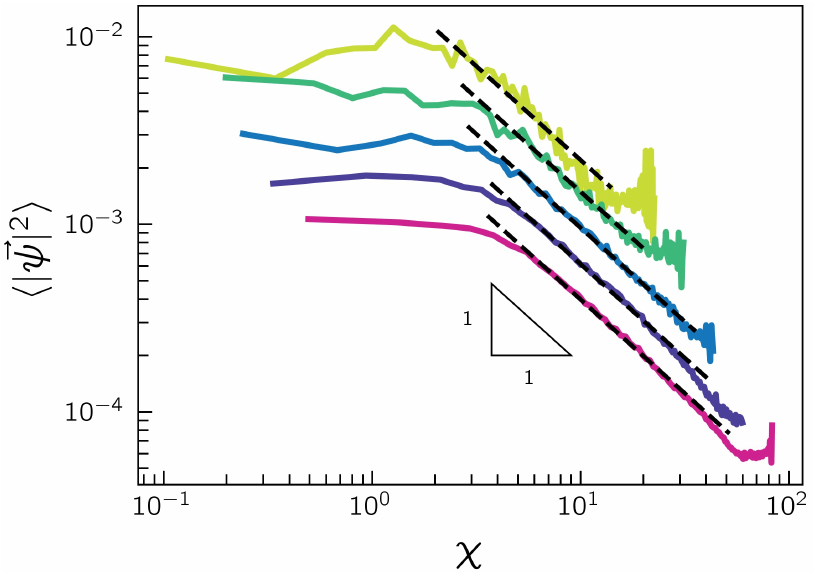}
\caption{\label{fig:append_mode} Spatial decay of vibrational magnitudes as a function of disatance from the exterior $\chi$ for systems with $v_0 = 0.5$ and varying system size. Color map is similar to that of Fig.~\ref{fig:pressure}. The black dashed lines show $\langle \vert \vec{\psi} \vert^2 \rangle \sim \chi^{-1}$ as a guide to the eye.}
\end{figure}

Similarly to the analysis of Sussman \textit{et.~al.}~in Ref.~\citenum{msussman_disordered_2015}, in this appendix, we study the spatial characteristics of low-frequency vibrational modes of the augmented Hessian. Fig.~\ref{fig:append_mode} shows the mean squared vibrational magnitude as a function of $\chi$ for systems with $v_0 = 0.5$ and $N \in \left\{256, 512, 1024, 2048, 4096 \right\}$. The average was taken over modes with $\omega \leq 0.4$ and over simulation duplicates. The results for systems with $v_0=0.25, 1.0$ are very similar.  Contrasting the results of Ref.~\citenum{msussman_disordered_2015}, we do not observe an exponential decay in the vibrational magnitude for any of the systems we examined. Rather, it appears that there is a plateau in $\vert \vec{\psi} \vert^2$ for small $\chi \lesssim 4.5$, followed by a $\langle \vert \vec{\psi} \vert^2 \rangle \sim \chi^{-1}$ power law decay. Further, we can identify a lengthscale $\chi^{*}$ associated with the onset of this $\chi^{-1}$ scaling. For the ensemble of vibrational modes studied here, $\chi^{*}\in(2.0,4.5)$ and increases monotonically with $N$. This analysis is consistent with the results we presented above for the vibrability of our packings, which reaches a plateau for large $\chi$.

\section{Higher-order interactions and role of large particle overlaps}\label{sec:appF_overlaps}

In Figs.~\ref{fig:vibrability}ac of the main text, it is clear that our prediction for the dominant contribution to vibrability near the center of the packings, $\Psi_{\text{QLM}}$, deviates from the measured vibrability plateau values, $\Psi_{\text{plat}}$, for systems with large values of $N$ and $v_0$. In this appendix, we study systems with a higher value of inter-particle interaction stiffness than the one shown in the main text. We expect that a higher value of stiffness $k$ will suppress the magnitude of overlaps and higher-order interactions, where more than two particles overlap each other. As these features are inherently present in two-dimensional packings and absent in our one-dimensional toy model for interaction pressure, we hypothesize that these effects contribute to disagreement with our scaling prediction and that increasing $k$ will therefore reduce the observed deviations.

\begin{figure}[ht]
\includegraphics[width = 0.95\linewidth]{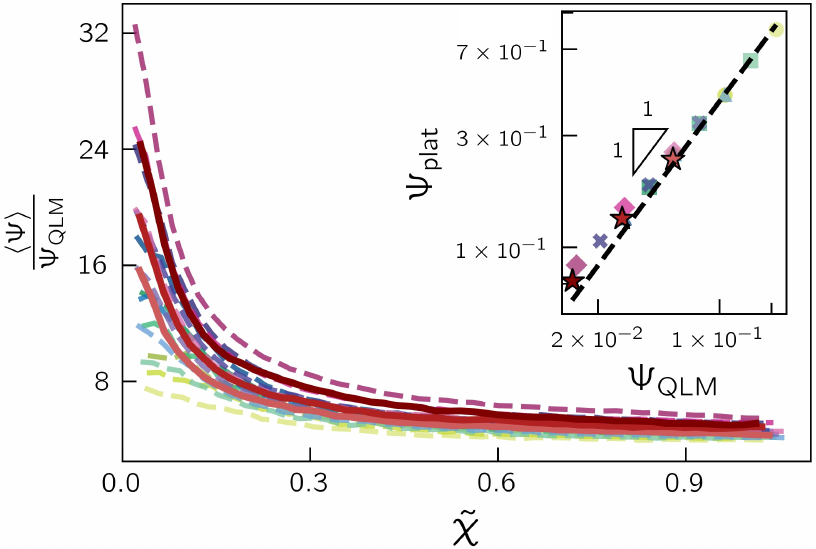}
\caption{\label{fig:append_interactions} Figs.~\ref{fig:vibrability}ac of the main text reproduced with additional rescaled vibrability data from a small (5-duplicate) ensemble of systems with $N=4096$, $v_0 = 0.5$, and $k=3000$. The original $k=1500$ data (dashed) has the same color map as in Fig.~\ref{fig:vibrability}a, and the additional $k=3000$ data (solid) is plotted in different shades of red corresponding to $v_0$ values ranging from $0.25$ (light, bottom) to $1.0$ (dark, top). The data in the inset has a similar color map, where star markers correspond to the $k=3000$ data. }
\end{figure}

Fig.~\ref{fig:append_interactions} shows the mean vibrability rescaled by $\Psi_{\text{QLM}}$ (see the main text and Appendix~\ref{sec:appD_barrier} above for details) as a function of $\tilde{\chi}$ for the ensemble of ($k=1500$) packings discussed in the main text (dashed lines) as well as a small ensemble with $N=4096$, $v_0 \in \left\{0.25, 0.5, 1.0 \right\}$, and $k=3000$ (solid lines) illustrating a better approximate collapse for large $\tilde{\chi}$. The inset to Fig.~\ref{fig:append_interactions} shows $\Psi_{\text{plat}}$ vs.~$\Psi_{\text{QLM}}$ for the same expanded dataset, confirming that indeed deviations from the scaling prediction are smaller in the systems with larger $k$.

\newpage 


\bibliography{bib_file}

\end{document}